\documentclass[useAMS,usenatbib,twocolumn]{mn2e}
\usepackage{amsmath}
\usepackage{graphicx}
\usepackage{color}
 \usepackage{multirow}

\def\simlt{\lower.5ex\hbox{$\; \buildrel < \over \sim \;$}}
\def\simgt{\lower.5ex\hbox{$\; \buildrel > \over \sim \;$}}
\def\simpt{\lower.5ex\hbox{$\; \buildrel \propto \over \sim \;$}}

\def\Mpc{\mbox{Mpc}}

\definecolor{mylabelcolor}{rgb}{0.5,1,1}

\title{Cosmological Information in the Gravitational
  Lensing of Pregalactic HI}

\date{\today}

\author[Metcalf \& White]{R. Benton Metcalf  
and S. D. M. White\\ Max Plank Institut f\"ur Astrophysics,  
Karl-Schwarzchild-Str. 1, 85741 Garching, Germany}

\begin{document}

\maketitle

\begin{abstract}
We study the constraints which the next generation of radio telescopes could place
on the nature of dark energy, dark matter and inflation by studying the
gravitational lensing of high redshift 21~cm emission, and we compare with the
constraints obtainable from wide-angle surveys of galaxy lensing.  If the
reionization epoch is effectively at $z\sim 8$ or later, very large amounts of
cosmological information will be accessible to telescopes like SKA and LOFAR.
We use simple characterizations of reionization history and of proposed
telescope designs to investigate how well the two-dimensional convergence
power spectrum, the three-dimensional matter power spectrum, the evolution of
the linear growth function, and the standard cosmological parameters can be
measured from radio data. The power spectra can be measured accurately over a
wide range of wavenumbers at $z\sim 2$, and the evolution in the cosmic energy
density can be probed from $z\sim 0.5$ to $z\sim 7$. This results in a
characterization of the shape of the power spectra (i.e. of the nature of dark
matter and of inflationary structure generation) which is potentially more
precise than that obtained from galaxy lensing surveys. On the other hand, the
dark energy parameters in their conventional parametrization ($\Omega_\Lambda,
w_o, w_a$) are somewhat less well constrained by feasible 21~cm lensing surveys than by
an all-sky galaxy lensing survey. This is because dark energy is felt
primarily at relatively low redshifts in this model; 21~cm surveys would be
more powerful than galaxy surveys for constraining models with ``early'' dark
energy.  Overall, the best constraints come from combining surveys of the two
types. This results in extremely tight constraints on dark matter and
inflation, and improves constraints on dark energy, as judged by the
standard figure of  merit, by more than an order of magnitude over
either survey alone. 
\end{abstract}


\section{introduction}
\label{sec:introduction}

The now widely accepted cold dark matter (CDM) model combines with
inflation-inspired scale-invariant primordial density fluctuations
to provide a consistent explanation for cosmic microwave background
(CMB) fluctuations, for type Ia supernova luminosity distances, for
the clustering of galaxies in redshift surveys, for the galaxy cluster
abundance and its evolution, and for the statistics both of weak
gravitational lensing and of Ly$\alpha$ forest absorption in quasar
spectra. This success comes at the cost of introducing several
mysterious and apparently {\it ad hoc} constituents.  Most of the
matter is supposed to be dark, a weakly interacting neutral particle
so far detected only through its gravitational effects.  All structure
is supposed to have originated during a very early period of
accelerated expansion driven by an inflaton field which has been
posited purely for this purpose.  Finally the energy density of the
Universe is apparently dominated today by a different and even more
unexpected field, dark energy, which is similarly accelerating the
present cosmic expansion. The nature of the dark energy and its
relation to the rest of physics are unknown. Detailed
measurements of the recent expansion history and of the corresponding
gravitationally driven growth of structure provide the only known
route to narrow down the possibilities.  The techniques mentioned
above can measure the expansion precisely out to $z\sim 4$ and the
growth of structure out to $z\sim 1.5$.  In this paper we investigate
a new technique that complements, refines and greatly increases the
precision of these methods by directly constraining both cosmic
expansion and structural growth out to $z\sim 6$.

The spin temperature of neutral hydrogen during and before the epoch
of reionization ($8 \simlt z \simlt 300$) fell out of thermal
equilibrium with the CMB radiation, resulting in the absorption and
emission of 21~cm radiation.  There has been a great deal of interest
in the prospect of detecting and mapping this radiation using
low-frequency radio telescopes, Several suitable instruments are now
under construction or in planning stages \citep[see][for an extensive
review]{astro-ph/0608032}.  This radiation provides an excellent
source for gravitational lensing studies.  Structure is expected in
the 21~cm emission down to arcsecond scales, and at each point on the
sky there will be $\sim 1000$ statistically independent regions,
separated in redshift (and thus frequency) that can in principle be
observed.  Gravitational lensing coherently distorts the 21~cm
brightness temperature maps at different frequencies.  This coherent
distortion can be distinguished from intrinsic structure in the HI gas
if enough independent redshifts are observed.  In this way a map of
the foreground density can be constructed
\citep{ZandZ2006,metcalf&white2006,HMW07,2007arXiv0710.1108L}.

Observing the 21~cm radiation at high redshift will be challenging. It
will require large arrays of radio telescopes working at low
frequencies ( $\sim 100$ to 200 MHz). At these frequencies foregrounds
-- terrestrial, galactic and extragalactic -- are very large and will
need to be subtracted by complex and as yet untested procedures
\citep{astro-ph/0608032}. This challenge is currently being addressed
by several observational teams (see section
\ref{sec:model-telescopes}).

In addition to the noise associated with mapping the brightness
temperature, the lensing signal has an additional {\it intrinsic
noise} component which comes from the unknown intrinsic structure of
the 21~cm brightness temperature distribution.  This noise cannot be
reduced by increasing the collecting area of the telescope, by
increasing the integration time or by improving the removal of
foregrounds.  \cite{metcalf&white2006} showed that if the
signal-to-noise in the brightness temperature map at each frequency is
greater than one, then the noise in the mass map will be close to the
intrinsic value.  Increasing the frequency resolution of the radio
observations increases the number of effectively independent regions
along the line-of-sight until the bandwidth becomes smaller than the
correlation length in the redshift direction of the brightness
temperature distribution.  If the bandwidth is matched to the
correlation length, the intrinsic noise is minimized.  The
correlation length in turn depends on beam size, and is smaller for
smaller beams.  Thus unlike galaxy lensing surveys, the intrinsic
noise for 21~cm lensing {\it decreases} with increasing telescope
resolution.  In practice, there is a trade-off because smaller
bandwidth means less flux, but this can be compensated by increasing
collecting area and/or integration time.  The noise is also affected
by the range of frequency (i.e. redshift) over which the 21~cm radiation
can be detected.  The low-frequency limit is set by the telescope
and/or the ability to subtract foregrounds.  The high-frequency limit
is typically set by the reionization of the universe, after which the
amount of intergalactic HI is small.

In \cite{HMW07} we simulated how well an idealized optimal telescope
would be able to map the 2 dimensional distribution of matter.  In
this paper, we study how well radio telescopes with specifications
similar to those of instruments currently being built or planned will
be able to constrain cosmological parameters, in particular those
related to the nature of dark matter, dark energy and inflation.

This paper is organized as follows.  The next two sections
introduce the formalism we use to study lensing in general and lensing
of the 21~cm radiation in particular.  In
section~\ref{sec:param-estim-dens} we develop a formalism for
extracting cosmological information and density maps from such lensing
data.  In section~\ref{sec:model-reionization} we discuss the relevant
aspects of reionization and the simple reionization model we use to
make quantitative predictions.  The telescope parameters used in our
predictions are given in section~\ref{sec:model-telescopes}.
Section~\ref{sec:2-dimens-inform} contains predictions for the noise
levels in various quantities.  A summary of our results and of
future prospects is given in the last section.

\section{lensing preliminaries}
\label{sec:lens-prel}

Gravitational lensing shifts the observed position of each point in
the image of a distant source.  Take the observed angular position on
the sky to be $\vec{\theta}$ and the position in the absence of
lensing to be $\vec{\beta}$. The first-order distortion in the image
is expressed by the derivatives of the mapping between these
angles. The distortion matrix is commonly decomposed into the
convergence $\kappa$ and two components of shear, $\bf \gamma$,
defined by
\begin{equation}
\left[ \frac{\partial \beta}{\partial \theta}\right] = \left(
\begin{array}{ccc}
1-\kappa +\gamma_1  & \gamma_2   \\
\gamma_2  &  1-\kappa-\gamma_1
\end{array}
\right).
\end{equation}

To lowest order and to an excellent approximation
\citep{2003ApJ...592..699V} the convergence is related directly to
the distribution of matter through
\begin{align}
\kappa\left(\vec{\theta},z_s\right) & = 
\frac{3}{4} H_o \Omega_m \int_0^\infty dz ~ \frac{(1+z)}{E(z)}
g\left(z,z_s \right) \delta\left(\vec{\theta},z\right) \label{eq:kappa_cont}\\
  & \simeq 
\frac{3}{4} H_o \Omega_m  \sum_i \delta\left(\vec{\theta},z_i\right) \int_{z_i-\delta
  z}^{z_i+\delta z} dz ~ \frac{(1+z)}{E(z)} g\left(z,z_s
\right)  \nonumber \\  
& =  \sum_i G(z_i,z_s) \delta\left(\vec{\theta},z_i\right) \label{eq:kappa_sum2}
\end{align}
with
\begin{equation}
g(z,z_s) = \int_{z}^\infty dz' ~ \eta\left(z',z_s\right)  
\frac{D(z,0)D(z',z)}{D(z',0)}.
\end{equation}
The weighting function for the source distance distribution,
$\eta(z)$, is normalized to unity.  $D(z',z)$ is the angular size
distance between the two redshifts and $\delta(\vec{x},z)$ is the
fractional density fluctuation at redshift $z$ and perpendicular
position $\vec{x}$.  The function
\begin{equation}
E(z)=\sqrt{\Omega_m(1+z)^3 + \Omega_\Lambda (1+z)^{3f(z)} +
  (1-\Omega_m-\Omega_\Lambda)(1+z)^2},
\end{equation}
where $\Omega_m$ and $\Omega_\Lambda$, are the present day densities
of matter and dark energy measured in units of the critical density.
The function describing the evolution of dark energy with redshift can
be written
\begin{equation}
f(z)=\frac{-1}{\ln(1+z)} \int^0_{-\ln(1+z)} \hspace{-0.6cm} \left[
  1+w(a) \right]  ~ d\ln a 
\end{equation}
where $w(a)$ is the equation of state parameter for the dark energy  -- the
ratio of the of its pressure to its density -- and $a=(1+z)^{-1}$ is the
scale parameter.  Where not otherwise mentioned, we will assume the
universe is flat, $\Omega_m+\Omega_\Lambda=1$, and $w(a)=-1$.
 
Equation~\eqref{eq:kappa_cont} shows that $\kappa$ can be interpreted
as a kind of projected dimensionless surface density.  For our
purposes it is convenient to express equation~(\ref{eq:kappa_sum2}) as
a matrix equation,
\begin{equation}
{\it \bf K} = {\rm \bf G} \boldsymbol{\delta},
\end{equation}
where the the components of ${\bf K}$ are the convergences running
over all position angles $\vec{\theta}$ and source redshifts, $z_s$.
The components of the vector $\boldsymbol{\delta}$ run over all
position angles and foreground redshifts $z_i$.  The matrix ${\rm \bf
G}$ is a function of most of the global cosmological parameters --
$\Omega_m$, $\Omega_\Lambda$, $w$, etc -- and is independent of
position on the sky.  The latter property makes these equations
equally valid when $\kappa$ and $\delta$ are transformed from angular
space to spherical harmonic space or to the $u$-$v$ plane where
interferometer observations are carried out.

When considering 21~cm lensing we will make the approximation
$\eta(z,z_s)=\delta^D(z-z_s)$, a Dirac delta function, for each band
observed corresponding to a frequency of $\nu = 1420
(1+z_s)^{-1}$~MHz. This is reasonable because angular size distances
vary little within a single band, but it is not true in the case of
galaxy lensing which we address in section~\ref{sec:cosm-param-1}.

\section{lensing of pregalactic HI}
\label{sec:lens-preg-hi}

Many convergence estimators are possible and can be expressed either
in real-space (which is more easily visualised) or in Fourier-space
(which is more easily related to interferometer observations in
visibility-space).  \cite{metcalf&white2006} gave a real-space
estimator for $\kappa$ that has some advantages, but to simplify the
present exposition we here use a Fourier-space estimator based on that
introduced by \cite{2002ApJ...574..566H} for the case of CMB lensing,
and extended by \cite{ZandZ2006} (and in the appendix of
\cite{metcalf&white2006}) to allow estimation of the 2-dimensional
$\kappa$ field from 3-dimensional 21~cm data.  Our treatment here
parallels this earlier work, although with some significant
differences in the source redshift weighting.

In the appendix of \cite{metcalf&white2006}) we applied the \cite{2002ApJ...574..566H} estimator for the Fourier transform of the convergence, $\kappa(\boldsymbol{\ell},\nu)$, to each frequency band and then combined them optimally to find a final estimate for $\kappa(\boldsymbol{\ell})$.  This has the disadvantage of requiring that an optimal bandwidth be found numerically which will be a function of both $|\boldsymbol{\ell}|$ and frequency.  An alternative method for combining frequencies is developed in \cite{ZandZ2006}.  They use the Fourier transform of $T(\ell,\nu)$ in the frequency direction as well as the angular directions and then find a convergence estimator of the form
\begin{align}\label{eq:ZZestimator}
\hat{\kappa}\left(\boldsymbol{\ell}\right) 
 = \int d^2\boldsymbol{\ell}' \sum_{k_1}\sum_{k_2}~
\chi(\boldsymbol{\ell}',\boldsymbol{\ell},k_1,k_2) 
T^*(\boldsymbol{\ell}',k_1) T^*(\boldsymbol{\ell}' - \boldsymbol{\ell},k_2)
\end{align}
where $\boldsymbol{\ell}$ is the Fourier dual of the angle and $k$ is the discrete Fourier mode in the radial direction.  The kernel $\chi(\boldsymbol{\ell}',\boldsymbol{\ell},k_1,k_2)$ is found by minimizing the noise while constraining the average to be the convergence.
This has the apparent advantage that the bandwidth does not come
into the calculation, but, in practice, an equally arbitrary frequency scale will need to be imposed.  In deriving the noise in their estimator \cite{ZandZ2006} make the
implicit assumption that the temperature field is statistically homogeneous in
frequency.  This assumption results in the Fourier modes being
uncorrelated which greatly simplifies the calculation.  In practice the
temperature field will be inhomogeneous because the noise will be
a strong function of frequency and because of the evolution of
structure and ionization.  As a result their estimator will not be optimal for a finite range in frequency only for $\Delta\nu/\nu \ll 1$.  This problem was avoided in our method.

In this paper we will adopt a hybrid method for estimating the convergence that avoids many of the shortcomings the previous two methods.  We use the \cite{ZandZ2006} estimator within frequency bands which are larger than our previous bandwidths and the minimum band width of the telescope, but small enough that no significant inhomogeneity in the noise and brightness temperature structure is expected.  This avoids having to find the optimal bandwidth.  We then linearly combine these $\kappa(\boldsymbol{\ell},\nu)$ estimates in optimal ways as described in detail in section~\ref{sec:param-estim-dens}.  We show that the correlations between convergence estimates at different frequencies is not significant if the bands widths are taken to be $\sim 1$~MHz.

The optimal kernel for estimator~\eqref{eq:ZZestimator} under the assumption of homogeneity and Gaussianity within the band is 
\begin{align}
\chi(\boldsymbol{\ell}',\boldsymbol{\ell},k_1,k_2,\nu) = \omega(\boldsymbol{\ell},\nu) |\boldsymbol{\ell}|^2 \delta_{k_1 k_2} 
~~~~~~~~~~~~~~~~~~~~~~~~~~~~~~~\nonumber \\
~~~~~~~~\times \frac{ \left[ \boldsymbol{\ell}\cdot \boldsymbol{\ell}'\, C_\nu(\ell',k_1) + \boldsymbol{\ell}\cdot (\boldsymbol{\ell}-\boldsymbol{\ell}')\, C_\nu(|\boldsymbol{\ell}'-\boldsymbol{\ell}|,k_1) \right] }{C^T_\nu(\ell',k_1) C^T_\nu(|\boldsymbol{\ell}'-\boldsymbol{\ell}|,k_1)} 
\end{align}
where $C^T_\nu(\ell,k)$ is the power spectrum of the actual
temperature, while $C_\nu(\ell,k)=C^T_\nu(\ell,k) +
C^N_\nu(\ell,k) $ is the observed power spectrum which
includes noise.  The normalization is
\begin{align}
\omega(\boldsymbol{\ell},\nu) = 
~~~~~~~~~~~~~~~~~~~~~~~~~~~~~~~~~~~~~~~~~~~~~~~~~~~~~~~~~  \nonumber \\
\frac{1}{2} \left[ \sum_k \int d^2{ \ell}' \frac{\left[
      \boldsymbol{\ell}\cdot \boldsymbol{\ell}'\,
      C_\nu(\ell',k) + \boldsymbol{\ell}\cdot
      (\boldsymbol{\ell}-\boldsymbol{\ell}')\,
      C_\nu(|\boldsymbol{\ell}'-\boldsymbol{\ell}|,k)
    \right]^2}{C^T_\nu(\ell',k)
    C^T_\nu(|\boldsymbol{\ell}'-\boldsymbol{\ell}|,k)}
\right]^{-1}.
\end{align}  



In the limit of an infinitely large beam the correlation between modes
is
\begin{eqnarray} \label{eq:shear_correlation}
\left\langle \hat{\kappa}\left({\bf L},k_1\right) \hat{\kappa}^*\left({\bf L}',k_2\right)  \right\rangle
 =  2(2\pi)^4 \delta^D\left({\bf L}-{\bf L}' \right) \delta_{k_1 k_2}
~~~~~~~~~~~  \nonumber \\
 \times \int d^2{ \ell}' ~\chi\left(\boldsymbol{\ell}',{\bf L},k_1\right)^2
 C^T_{\nu}(\ell',k_1) C^T_{\nu}(|\boldsymbol{\ell}'-{\bf L}|,k_1)
  \\  \label{eq:shear_correlation2}
 =  (2\pi)^2 \delta^D\left({\bf L}-{\bf L}' \right) \delta_{k_1 k_2}
 N^{\hat{\kappa}}(L,\nu) 
\end{eqnarray}
To simplify analysis, it has here been assumed that the temperature
(i.e. the 21~cm emissivity) is Gaussian distributed so that the fourth
moment can be written as products of second moments.  The actual temperature
distribution will be non-Gaussian because of non-uniform spin
temperature, peculiar velocities, nonlinear structure
formation, and non-uniform ionization.  The importance of nonlinear
structure formation in this context has been highlighted by
\cite{2007arXiv0710.1108L}.  Some of these non-Gaussian effects, such
as non-uniform ionization, cannot be usefully quantified at this
time.  The Gaussian approximation should be valid at least before
significant reionization occurs and is probably good while the reionized
regions are significantly smaller than the resolution of the telescope.

The finite telescope beam will cause correlations in the noise, $\langle
\kappa(\boldsymbol{\ell})
\kappa^*(\boldsymbol{\ell}+\delta\boldsymbol{\ell})\rangle \neq
0$, when $\delta\ell \simlt 2\pi \sigma_u$ where $\sigma_u$ is the
width of the beam in $u$-$v$ space \citep{metcalf&white2006}.  These
correlations can be taken into account, but for simplicity they will
not be considered here.

The noise in $\hat{\kappa}(\ell,\nu)$ within one frequency band  is
\begin{align}\label{N_L_nocorr}
N^{\hat{\kappa}}(\ell,\nu) 
 = \frac{(2\pi)^2}{2}
 ~~~~~~~~~~~~~~~~~~~~~~~~~~~~~~~~~~~~~~~~~~~~~~~~~~~~~~~
 \nonumber \\ \times 
\left[\sum_k \int d^2\ell'~ \frac{ \left[ \boldsymbol{\ell}\cdot \boldsymbol{\ell}'\, C_\nu(\ell',k) + \boldsymbol{\ell}\cdot (\boldsymbol{\ell}-\boldsymbol{\ell}')\, C_\nu(|\boldsymbol{\ell}'-\boldsymbol{\ell}|,k) \right]^2 }{C^T_\nu(\ell',k) C^T_\nu(|\boldsymbol{\ell}'-\boldsymbol{\ell}|,k)} \right]^{-1}
\end{align}
Because the estimator~\eqref{eq:ZZestimator} is a sum over all the
observed pairs of visibilities, it will (by the central limit theorem)
be close to Gaussian distributed even though it is quadratic in the
visibilities.  In the remainder of this paper this property will be
assumed.  The intrinsic noise limit corresponds to the case where
$C_\nu(\ell) = C^T_\nu(\ell)$.

In real data the temperature distribution will not be Gaussian, the
foregrounds will not be perfectly subtracted and will produce spurious
frequency correlations, there will be holes in the area surveyed
around bright point sources, there will be a finite and irregular
beam, and the coverage of the $u$-$v$ plane will not be complete.  It is also true that the convergence estimator was derived in the weak lensing limit ($\kappa \ll 1$) which will not be valid in all regions of the sky.  All
these complications make it unclear at the present time what estimator
will be optimal for a real experiment.  We 
nevertheless believe that the above relatively simple assumptions
should give a good indication of what can be expected.

\section{parameter estimation and density mapping}
\label{sec:param-estim-dens}

The noise in $\hat{\kappa}(\ell,z)$ for different bands and different $\ell$
will, to a good approximation, be statistically independent. 
The data vector will be defined as
\begin{align}
  \label{eq:datavector}
  {\bf D} &= \hat{\bf K}-{\bf K} \\
  &= \hat{\bf K}-{\bf G} \boldsymbol{\delta}
\end{align}
where the components run over all the combinations of $z_i$ and $\ell$
that are measured.

In this case the log of the likelihood function can be written
\begin{equation}
  \label{eq:likelihood}
 \ln {\mathcal L} = - \frac{1}{2}  {\bf D}^\dag {\rm\bf N}^{-1} {\bf
     D}  - \frac{1}{2} |{\rm\bf N}| -  H .
\end{equation}
An additional function $H$ has been added to represent a prior
distribution on the parameters or, in the context of density
reconstruction, a regularization.  As written so far the free
parameters include all the cosmological parameters and all the
foreground densities in each redshift bin.  The noise covariance
matrix is approximately
\begin{align}\label{eq:diagonal-noise}
 {\rm\bf N}_{ij} & =  I_{ij} N^{\hat{\kappa}}(\ell_i,\nu_i) \\\label{eq:diagonal-noise2}
& \simeq \delta_{ij}  I(\ell_i,\nu_i) N^{\hat{\kappa}}(\ell_i,\nu_i).
\end{align}
 The matrix $I_{\nu\nu'}({\bf L})$ has been introduced to express possible 
 cross-correlations between frequency bands. It will be normalized so that 
 $I_{\nu\nu}({\bf L})=1$.  The second line further simplifies this by assuming that the noise is equal in frequency bands that have significant correlation between them.  In is case the $I_{\nu\nu'}({\bf L})$ matrix can be made into a factor, $I(\ell,\nu)= \sum_{\nu'} I_{\nu\nu'}(\ell)$.  In our studies here $I_{\nu\nu'}({\bf L}) \simeq \delta_{\nu\nu'}$ or $I(\ell,\nu) \simeq 1$ since the correlations between the wide bands used ($\sim 1$~MHz) are small, but it is possible that foreground subtraction in particular might introduce significant correlations between frequency bins. Foreground subtraction will not be discussed in detail here, but we will retain the off-diagonal elements in our formalism.
There will also be some off-diagonal elements to ${\bf N}_{ij}$ between different $\ell$ values caused by the finite beam of the telescope. Thses could be incorporated in a future analysis.

The maximum likelihood estimate for any parameter can be found by
maximizing (\ref{eq:likelihood}) with respect to that parameter.  The
error in this estimator is often forecasted using the Fisher matrix
defined as
\begin{equation}\label{eq:likely_diag}
 {\rm\bf F}_{ij} = - \left\langle \frac{\partial^2
     \ln\mathcal{L}}{\partial p_i \partial p_j} \right\rangle.
\end{equation} 
The expected error in the parameter $p_a$, marginalized over all other
parameters, is $\sigma_a^2 \simeq \left({\rm\bf
F}^{-1}\right)_{aa}$. The unmarginalized error estimate (the error
when all other parameters are held fixed) is $\left( {\rm\bf
F}_{aa}\right)^{-1}$.

Often the parameters of physical interest have highly correlated
noise, quantified by the off-diagonal elements in ${\bf F}$.  As a
result $\sigma_a$ can be a deceptive measure of how well the data
constrains the parameter set as a whole. A way to mitigate this
is to find the transformation that diagonalizes ${\bf F}$
\begin{equation}\label{eq:fisher_diag}
 {\rm\bf F} = {\bf V}^\dag \boldsymbol{\lambda} {\bf V}.
 \end{equation}
This defines linear combinations of the parameters ${\bf
  \hat{p}}={\bf V}{\bf p}$ that are uncorrelated and have variances
$\lambda_{aa}^{-1}$.  This will be used in section~\ref{sec:tomogr-inform}.

\subsection{Tomography}
\label{sec:tomography}

If one is primarily interested in reconstructing the cosmic mass
density distribution, the background cosmology can be held fixed and
$\mathcal{ L}$ can be maximized with respect to the pixelized
foreground density.  This can be done in 2-D by approximating
$\kappa$ as independent of $\nu$, in which case the solution that
maximizes \eqref{eq:likelihood} is
\begin{equation}
\kappa(\vec{\ell}) =\frac{ \sum_{\nu\nu'} {\bf N}_{\nu\nu'}^{-1}(\ell)
  ~\hat{\kappa}(\vec{\ell},\nu) }{\sum_{\nu\nu'} {\bf N}_{\nu\nu'}^{-1}(\ell)},
\end{equation}
with corresponding noise,
\begin{align}\label{eq:2Dnoise}
N(\ell) & =  \frac{1}{\sum_{\nu\nu'} {\bf N}_{\nu\nu'}^{-1}(\ell)} \\
 & \simeq  \frac{1}{\sum_{\nu} \left[ I(\ell,\nu) N(\ell,\nu) \right]^{-1}} ,
\end{align}
which is uncorrelated between $\vec{\ell}$ values that are separated
by more than the resolution of the telescope.  Here no regularizing
function is used ($H=0$).  This is the result derived by
\cite{metcalf&white2006}.

A 3-D density reconstruction can be obtained by maximizing
(\ref{eq:likelihood}) with respect to the components of
$\boldsymbol{\delta}$.  For $H=0$, the solution, after some algebra and taking
into account that ${\bf N}$ is symmetric, is
\begin{equation} \label{delta_estimator}
\boldsymbol{\delta} = \left[ {\bf G}^\dag{\bf N}^{-1} {\bf G}
\right]^{-1} {\bf G}^\dag {\bf N}^{-1} \hat{\bf K}.
\end{equation}
Essentially the same result was obtained by \cite{2002PhRvD..66f3506H}
for density reconstruction using galaxy lensing data in real-space.
This requires that the matrix in braces be invertible.  In the special
case where ${\bf G}$ is invertible the estimator is simply
$\boldsymbol{\delta} = {\bf G}^{-1} \hat{\bf K}$.  The noise
covariance matrix for the reconstruction (\ref{delta_estimator}) is
\begin{equation} \label{delta_noise}
{\bf N}_\delta= \left[ {\bf G}^\dag {\bf N}^{-1} {\bf G}\right]^{-1}.
\end{equation}
This noise will be highly correlated between $\delta$'s.

The regularization function in (\ref{eq:likelihood}) can be used to
improve the density reconstruction at the expense of making some
assumptions about the statistical properties of the underlying density
distribution.  One choice is to assume that this
distribution is Gaussian, in which case
\begin{align}
H & = \frac{1}{2} \boldsymbol{\delta}^\dag {\bf C}^{-1}_\delta \boldsymbol{\delta}
\\
& = \frac{1}{2} \sum_{i\ell} \frac{\delta(z_i,\vec{\ell})^2}{C_\delta(z_i,\ell)},
\end{align}
where $C_\delta(z,\ell)$ is the angular power spectrum of mass within
the redshift bin labeled by $z_i$.  This is equivalent to a Wiener
filter.  The optimal estimator~(\ref{delta_estimator}) and the noise
(\ref{delta_noise}) will be modified in this case, but can be derived
in the same way.  In addition to smoothing the noisy map,
regularization provides a well behaved way to deal with missing
telescope baselines and with holes in coverage due to foreground
sources.  Other regularization schemes could include an entropic prior
or filters designed to emphasize localized mass lumps
\citep{2002PhRvD..66f3506H}.

\subsection{Cosmological parameter estimation}
\label{sec:cosm-param}

Estimates of cosmological parameters and statistical information about
the underlying mass density distribution can be extracted from 21~cm
data without constructing density maps.  The actual distribution of
matter can be marginalized over assuming a suitable prior.  This
procedure can give surprisingly precise results, even when noise
levels are far too high for a meaningful density reconstruction to be
possible.

According to the standard model of structure formation, the components
of ${\bf K}$ (i.e. $\kappa(\boldsymbol{\ell},\nu)$) will be normally
distributed for $\ell$ less than several thousand
\citep{2004MNRAS.348..897T}.  In this case a prior $H=\frac{1}{2} {\bf
K}^\dag {\bf C}_\kappa^{-1} {\bf K}$ can be used and the likelihood
function can be integrated over all components of ${\bf K}$ --
i.e. over all possible convergence maps.  ${\bf C}_\kappa$ here is the
(cross-)power spectrum of the convergence for two different source
redshifts,
\begin{equation}
\left[{\bf C}_\kappa\right]_{ij} = \left\langle
\kappa(\vec{\ell},z_i) \kappa(\vec{\ell},z_j) \right\rangle.
\end{equation}
This can be calculated using expression~(\ref{eq:kappa_cont}) and a
model for the matter power spectrum.  The resulting likelihood
function has the same form as (\ref{eq:likelihood}) 
\begin{equation}
  \label{eq:likelihood2}
 \ln {\mathcal L} = - \frac{1}{2}  \hat{\bf K}^\dag {\rm\bf C}^{-1} \hat{\bf
     K}  - \frac{1}{2} |{\rm\bf C}|, 
\end{equation}
where the noise matrix, ${\bf N}$, is now replaced with the covariance
matrix,
\begin{equation}
{\bf C} = {\bf N} + {\bf C}_\kappa.
\end{equation}

Since $\ell$-modes separated by more than the resolution of the
telescope will not be correlated, we can break the likelihood function
up into factors representing each resolved region in $\ell$-space
\citep{metcalf&white2006}.  The result is that there are $\sim
(2\ell+1)f_{\rm sky}$ independent measured modes for each value of
$\ell$, where $f_{\rm sky}$ is the fraction of sky surveyed.  The
Fisher matrix can then be further simplified to the widely used form,
\begin{equation}\label{eq:fisher}
{\bf F}_{ab} = \frac{1}{2} \sum^{\ell_{\rm max}}_{\ell=\ell_{\rm min}} (2\ell +1) f_{\rm sky} {\rm tr}\left[ {\bf C}^{-1} {\bf C},_a {\bf C}^{-1} {\bf C},_b \right]~.
\end{equation}

A quantity of particular interest that will be calculated later in this
paper is the noise variance in the 2-D convergence power spectrum estimate.
This is easily derived from (\ref{eq:fisher}) as
\begin{equation}\label{eq:DeltaC}
 \Delta C_\kappa(\ell) = \sqrt{\frac{2}{(2\ell+1)f_{\rm sky}} } \, \left[ C_\kappa(\ell) + N(\ell)\right],
\end{equation}
where $N(\ell)$ is given by (\ref{eq:2Dnoise}).  The first term
represents the sample or cosmic variance, and the second the noise in
the $\kappa$ estimate itself.  To increase the signal-to-noise the
average power (known as the band-power) can be estimated within a bin
of width $\Delta\ell$ larger than the resolution limit of the survey,
$\delta\ell \sim f_{\rm sky}^{-1/2}$.  The noise in the band-power is
the above divided by the square root of the number of independent
measurements within the band,
\begin{equation} \label{eq:DeltaCbinned}
 \Delta C_\kappa\left(\ell,\Delta\ell\right) \simeq \frac{\Delta
   C_\kappa(\ell)}{f_{\rm sky}^{1/4} \Delta\ell^{1/2}}.
\end{equation}

\section{reionization model}
\label{sec:model-reionization}

The fluctuations in the brightness temperature depend on the spin
temperature, $T_s$, the ionization fraction, $x_{\rm H}$ and the
density of HI through
\begin{align}
\delta T_b \simeq 24  (1+\delta_b) x_{\rm H}\left( \frac{T_s- T_{\rm
CMB}}{T_s}\right)\left(\frac{\Omega_b h^2}{0.02}\right) \nonumber \\ 
\times \left( \frac{0.15}{\Omega_m h^2}\frac{1+z}{10} \right)^{1/2} 
\mbox{ mK} 
\end{align}
\citep{1959ApJ...129..536F,1997ApJ...475..429M}.  As is commonly done,
we will assume that the spin temperature is much greater than the CMB
temperature.  This leaves fluctuations in $x_H$, and the baryon
density $\delta_b = (\rho_b - \overline{\rho}_b)/\overline{\rho}_b$ as
the sources of brightness fluctuations.  We will make the simplifying
assumption that $x_{\rm H}=1$ until the universe is very rapidly and
uniformly reionized at a redshift of $z_{\rm reion}$.  Realistically,
the reionization process will be inhomogeneous and may extend over a
significant redshift range.  This will increase
$C_\nu(\ell)$ by perhaps a factor of 10 on scales larger
the characteristic size of the ionized bubbles
\citep{2004ApJ...608..622Z} and thus might be expected to reduce the
noise in the lensing map, $\hat{\kappa}(\vec{\theta})$, significantly.
However, the noise, (\ref{N_L_nocorr}), has been derived
assuming that the fourth order statistics of $\delta T_b$ can be
approximated by the values appropriate for a Gaussian random field.  If
this approximation holds, the lensing noise will indeed be reduced
during reionization, but this is uncertain, since the temperature
distribution will clearly not be Gaussian at this time, especially
when the neutral fraction is low.  Incorporating a more realistic
ionization history is difficult and requires further work.  A
definitive resolution of these uncertainties will not, in any case, be
possible until the observations are taken.

With real data one will not need to rely on an assumption of
Gaussianity.  It will be possible to derive the uncertainties by
bootstrap techniques, since the higher order statistics of the
temperature variations can be estimated directly from the data
themselves.

\section{model telescopes}\label{sec:model-telescopes}

In order to forecast the capabilities of future instruments, we will
present results assuming several sets of telescope parameters. These
are intended to represent the characteristics of facilities currently
under construction or in an advanced stage of planning.

\begin{table*}
\begin{minipage}{14cm}
\caption{Parameters for telescope models.}
\begin{tabular}{c|ccccccc}
model  & $D$ (km)      & $f_{\rm cover}$ & $t_o$ (days) & $\nu_{\rm
  min}$ (MHz) & $\Delta\nu$ (MHz) & $z_{\rm reion}$ & reionization \\
\hline
LOFARII-1yr & 2 & 0.016 & 30 & 110 & 1 & 7 & instantaneous\\
LOFARII-3yr & 2 & 0.016 & 90 & 110 & 1 & 7 & instantaneous\\
SKA-1yr   & 6 & 0.020  & 30 & 100 & 1 & 7 & instantaneous\\
SKA-3yr   & 6 & 0.020  & 90 & 100 & 1 & 7 & instantaneous\\
SKA-3yrA   & 6 & 0.020  & 90 & 100 & 1 & 10 & instantaneous\\
\hline
\end{tabular}
\label{table:params}
\end{minipage}
\end{table*}

The noise in each visibility measurement will have a thermal component
and a component resulting from imperfect foreground subtraction.  Here
we model only the thermal component.  If the telescopes in the array
are uniformly distributed on the ground, the average integration time
for each baseline will be the same and the power spectrum of the noise
will be
\begin{align}\label{eq:C_noise}
 C_\ell^N = \frac{2\pi}{ \Delta\nu t_o} \left(  \frac{T_{\rm sys} \lambda}{ f_{\rm cover} D_{\rm tel}} \right)^2 = \frac{(2\pi)^3 T_{\rm sys}^2}{\Delta\nu t_o f_{\rm cover}^2 \ell_{\rm max}(\nu)^2}, 
\end{align}
\citep{2004ApJ...608..622Z,2005ApJ...619..678M,2006ApJ...653..815M}
where $T_{\rm sys}$ is the system temperature, $\Delta\nu$ is the
bandwidth, $t_o$ is the total observation time, $D_{\rm tel}$ is the
diameter of the array and $\ell_{\rm max}(\lambda)=2\pi D_{\rm
tel}/\lambda$ is the highest multipole that can be measured by the
array, as set by the largest baselines.  $f_{\rm cover}$ is the total
collecting area of the telescopes divided by $\pi (D_{\rm tel}/2)^2$,
the aperture covering fraction.  Other telescope configurations are
possible which would result in the noise being unequally distributed
in $\ell$, but here we will consider only this uniform configuration.
At the relevant frequencies, the overall system temperature is
expected to be dominated by galactic synchrotron radiation.  We will
approximate the brightness temperature of this foreground as $T_{\rm
sky}=180 \mbox{ K}(\nu/180\mbox{ MHz})^{-2.6}$, as appropriate for
regions well away from the Galactic Plane \citep{astro-ph/0608032}.
This results in larger effective noise for higher redshift
measurements of the 21~cm emission.

Several relevant telescopes are currently proposed or under
construction.  The 21 Centimeter Array (21CMA, formerly known as
PAST)\footnote{21cma.bao.ac.cn/} has $f_{\rm cover} \sim 0.01$ and
$\ell_{\rm max} \sim 10^3$ giving it a resolution of about 10~arcmin.
The Mileura Widefield Array (MWA) Low Frequency Demonstrator
(LFD)\footnote{www.haystack.mit.edu/ast/arrays/mwa/} will operate in
the 80-300~MHz range with $D_{\rm tel}\simeq 1.5$~km and $f_{\rm
cover} \sim 0.005$.  For LOFAR (the Low Frequency
Array)\footnote{www.lofar.org} the core array was originally planned to have $f_{\rm cover}
\sim 0.016$ and $D_{\rm tel} \sim 2$~km.  LOFAR's extended baselines,
reaching to 350~km and possibly farther, will not be useful for
high redshift 21~cm observations because of the small $f_{\rm cover}$,
although they will be helpful for foreground subtraction.  It is
anticipated that LOFAR will be able to detect 21~cm emission out to
redshift $z\simeq 11.5$, but sensitivity limitations will make
mapping very difficult.  Recently budget constraints have reduced the 
initial collecting of the telescope core by about one third.  We will use approximations to these original  plans for our forecasting and call our mock telescope LOFARII to indicate that it is larger than the first stage of LOFAR being constructed.
Plans for the SKA (the Square Kilometer
Array)\footnote{www.skatelescope.org/} have not been finalized, but it
is expected to have $f_{\rm cover} \sim 0.02$ over a diameter of
$\sim 6$~km ($\ell_{\rm max} \sim 10^4$) and sparse coverage extending
out to 1,000-3,000~km.  The lowest frequency currently anticipated is
$\sim 100$~MHz which corresponds to $z\sim 13$.  It is expected that
the core will be able to map the 21~cm emission with a resolution of
$\Delta\theta \sim 1\mbox{ arcmin}$. For reference, one arcminute
(fwhm) corresponds to baselines of 5.8~km at $z=7$ and 11~km at
$z=15$.  

The various sets of telescope and survey parameters we use below are
listed in Table~\ref{table:params}. We adopt values appropriate to one
or three seasons of observing with the core arrays of LOFARII and SKA, and
we assume that reionization occurs instantaneously at $z=7$ or (for
one SKA case) at $z=10$.

\section{forecasts}
\label{sec:2-dimens-inform}

In the following subsections the formalism of sections
\ref{sec:lens-prel} through \ref{sec:param-estim-dens} and the models
of sections~\ref{sec:model-reionization} and
\ref{sec:model-telescopes} are used to forecast how well future
projects will be able to estimate various quantities of interest.  As
background cosmological model we choose $\Omega_{\rm matter}=0.3$
(matter density), $\Omega_\Lambda=0.7$ (cosmological constant),
$H_o=70$ (Hubble parameter), $\sigma_8=0.75$ (normalization of the
matter power spectrum) and $\Omega_b=0.03$ (baryon density).  Except
in section~\ref{sec:matt-power-spectr} the (nonlinear) matter power
spectrum is calculated using the method of \cite{peac96}.
In the following we will make the approximation $I(L,\nu)=1$ which will be further justified.

\subsection{Convergence power spectrum}
\label{sec:conv-power-spectr}

\begin{figure} \rotatebox{90}{
\includegraphics[width=6.0cm]{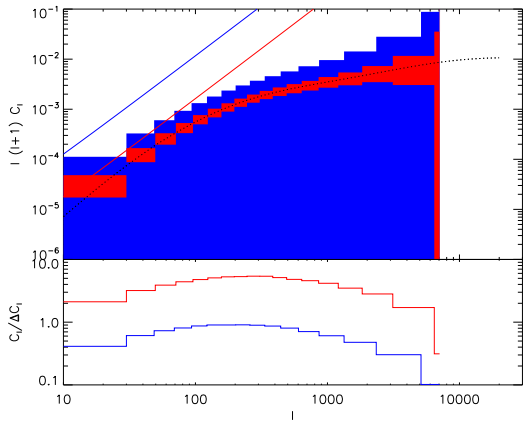}} 
\caption{Forecasts of the 1$\sigma$ uncertainties in estimates of the
convergence power spectrum for our LOFARII-1yr (blue) and LOFARII-3yr
(red) parameter sets.  In the upper panel the solid straight lines give the noise per
mode, while the dotted curve is the underlying model power spectrum.
We have here assumed that 25\% of the sky has been observed (the noise
scales as $f_{\rm   sky}^{-1/2}$). The band powers are for bins in 
$\ell$ chosen according to formula (\ref{eq:l_bins}) with
$f_{\rm bin}=10$. The signal-to-noise ratio for the banned power is
given in the lower panel. } 
\label{fig:Ckappa_LOFAR}
\end{figure}

\begin{figure}
\rotatebox{90}{
\includegraphics[width=6.0cm]{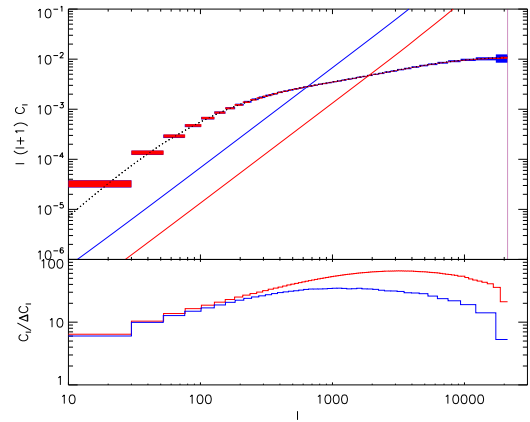}}  
\caption{Forecasts of the 1$\sigma$ uncertainties in estimates of the
convergence power spectrum for our SKA-1yr (blue) and SKA-3yr (red)
parameter sets. The band powers are for bins in $\ell$ chosen
according to formula (\ref{eq:l_bins}) with $f_{\rm bin}=5$. All other
parameters are the same as in Figure~\ref{fig:Ckappa_LOFAR}. Note that
below about $\ell\sim 10^3$ the uncertainties are dominated by cosmic
variance and do not improve for the longer observing time.} 
\label{fig:Ckappa_SKA} 
\end{figure}

\begin{figure}
\rotatebox{90}{
\includegraphics[width=6.0cm]{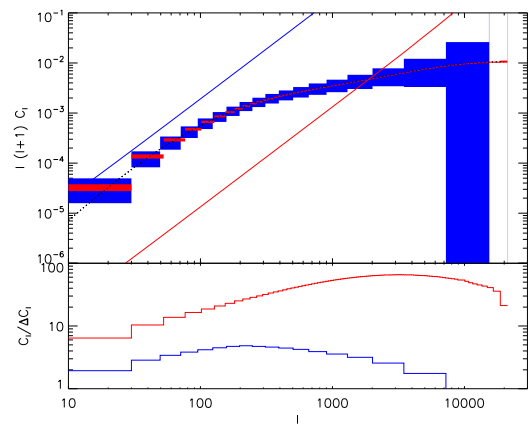}}  

\caption{Forecasts of the 1$\sigma$ uncertainties in estimates of the
convergence power spectrum for our SKA-3yr (red) and SKA-3yrA (blue)
parameter sets. The band powers are for bins in $\ell$ chosen
according to formula (\ref{eq:l_bins}) with $f_{\rm bin}=10$.  All
other parameters are the same as in Figure~\ref{fig:Ckappa_LOFAR}.
The difference between these two examples is the reionization
redshift, $z_{\rm reion}=7$ for SKA-3yr and $z_{\rm reion}=10$ for
SKA-3yrA.}  
\label{fig:Ckappa_SKAz} 
\end{figure}

\begin{figure}
\rotatebox{90}{
\includegraphics[width=6.0cm]{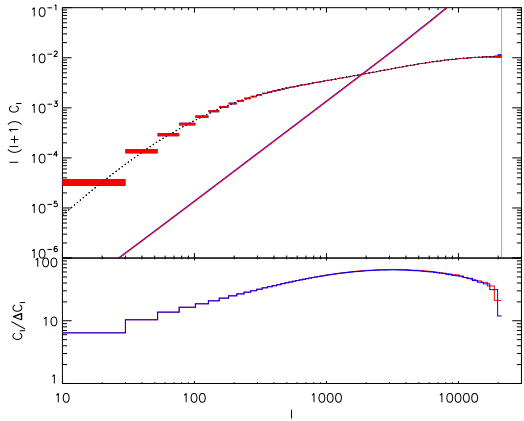}}  
\caption{An illustration that increasing the frequency bandwidth of
the observations results in increased noise in the $\kappa$
estimate. Red shows results for our SKA-3yr parameters which adopt a
bandwidth of 1~MHz, while blue adopts the same parameters except
that the bandwidth is taken to be 5~MHz. The band powers here are
for $\ell$ bins with $f_{\rm bin}=10$.  All other parameters are the
same as in Figure~\ref{fig:Ckappa_LOFAR}.}
\label{fig:Ckappa_SKAbandwidth} 
\end{figure}

The first objective of a 21~cm lensing survey might be to measure the
two-point statistics of the convergence field
$\kappa(\vec{\ell},z_s)$, averaged over $z_s$.  The expected error in
the binned power spectrum $C_\kappa$ is given by
equations~(\ref{eq:DeltaC}) and (\ref{eq:DeltaCbinned}).  For purely
aesthetic reasons we choose the $\ell$-space bin-widths to be
\begin{equation}
  \label{eq:l_bins}
  \Delta\ell = \frac{f_{\rm bin}}{f_{\rm sky}} \left[
    \frac{C_\kappa(\ell=10)}{\Delta C_\kappa(\ell=10)}\frac{\Delta C_\kappa(\ell)}{ C_\kappa(\ell)} \right] \left(
    \frac{\ell}{10} \right)^{3/5}.
\end{equation}
Here, the bins at $\ell=10$ are $f_{\rm bin}$ times the telescope
resolution, where the factors $f_{\rm bin}$ used for figures~\ref{fig:Ckappa_LOFAR} through \ref{fig:Ckappa_SKAbandwidth} are given in the captions.

Figure~\ref{fig:Ckappa_LOFAR} shows forecasts for a LOFARII-like
telescope with observing times of 30~days (approximately one season)
and 90~days.  Reionization is assumed to occur at $z=7$.  It can be
seen that the noise per mode is nowhere smaller than the
signal per mode, even after three seasons.  Thus, LOFARII will be unable
to make a high-fidelity map of the density distribution unless
reionization is very different than assumed here.  Despite this, after
three seasons the band-averaged convergence power spectrum can be
recovered with good signal-to-noise over a wide range of
$\ell$-values.

Figure~\ref{fig:Ckappa_SKA} shows similar forecasts for an SKA-like
telescope.  Here the noise per mode is below the signal for $\ell
\simlt 700$ after 30 days observing, and for $\ell \simlt 2000$ after
90~days, so mapping the convergence on these scales should be
possible.  This agrees with the conclusions in
\cite{metcalf&white2006}, although we used a different criterion for
judging a ``good'' map in that paper. In cases such as this, where the
noise per mode is smaller than the expected power in the signal, the
uncertainty in an estimate of the power spectrum is dominated by
cosmic variance, and the only way to increase the precision of the
measurement is to survey a larger fraction of the sky.  This can be
seen in figure~\ref{fig:Ckappa_SKA} where the noise becomes
independent of observing time at small $\ell$.  With the parameters
assumed here, an SKA-like telescope will clearly be able to make an
excellent estimate of the convergence power spectrum.

Figure~\ref{fig:Ckappa_SKAz} illustrates how uncertainties in the
power spectrum estimate depend on the redshift of reionization.  We
compare results for instantaneous reionization at redshifts of 7 and
10 for 3 years of observation with an SKA-like telescope.  When
$z_{\rm reion}$ is high, there are fewer independent sources of 21~cm
emission in the observable frequency range and the noise goes up
significantly.  Nevertheless, using only the emission between $z=10$
and 13, a good convergence power spectrum can still be recovered,
although mapping the dark matter distribution would no longer be
possible.  A telescope that goes to lower frequencies could tolerate
higher reionization redshifts without such performance degradation. An
example is the MWA, although it currently has insufficient collecting
area for this project. A LOFAR-like telescope would be unable to
obtain a useful power spectrum estimate for $z_{\rm reion} \simgt 9$.

Figure~\ref{fig:Ckappa_SKAbandwidth} illustrates how the uncertainties
depend on the bandwidth.  Increasing the bandwidth by a factor of 5 makes little difference to the results.  It is assumed here that the  statistical properties of the noise and source 21~cm radiation are constant within each band.  This result indicates that the correlations between frequency bands when $\delta\nu = 1$~MHz are not of relevance since they are fully incorporated when the bandwidth is increased.  As discussed before, the estimator is optimized for the case where the statistical properties of the noise and source 21~cm radiation are constant across each band and
so it is better to use a smaller bandwidth where this assumption is better justified.

The lensing map of the cosmic mass distribution can have good fidelity while  the temperature map of high redshift 21cm emission is noise-dominated on
the same scale.  This somewhat counter-intuitive situation reflects
the fact that it is better to have more independent redshift slices at
low signal-to-noise than to make high-quality images of the brightness
temperature in a small number of channels.

\subsection{Matter power spectrum}
\label{sec:matt-power-spectr}

\begin{figure}
\rotatebox{90}{
\includegraphics[width=6.0cm]{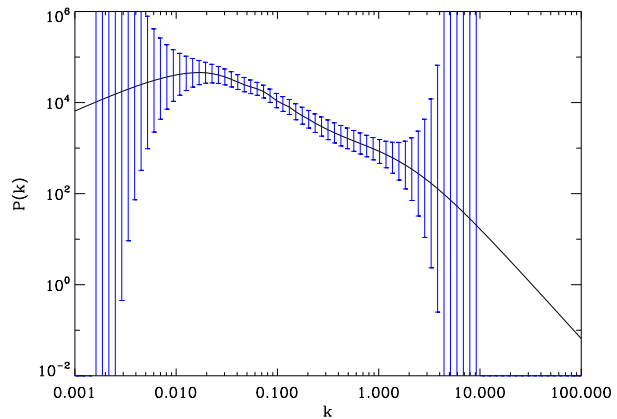}}  
\caption{Forecast uncertainties in a reconstruction of the matter
power spectrum from lensing of high-redshift 21~cm emission for
observations with our LOFARII-3yr parameter set. We have taken 7.93
points per decade in the model power spectrum and have used no
tomographic information. We assume $f_{\rm sky}=0.25$.}
\label{fig:pofk_LOFAR} 
\end{figure}

\begin{figure}
\rotatebox{90}{
\includegraphics[width=6.0cm]{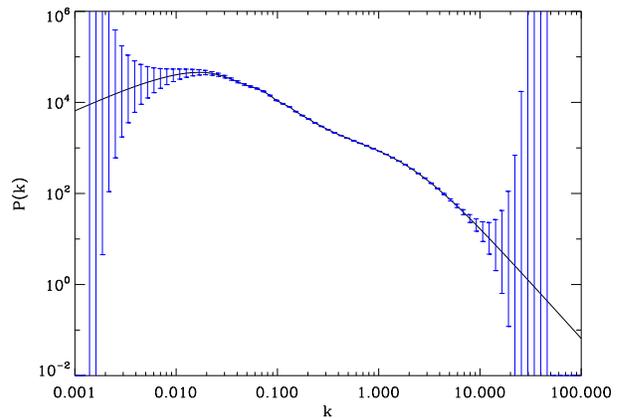}}  
\caption{Forecast uncertainties in a reconstruction of the matter
power spectrum from lensing of high-redshift 21~cm emission for
observations with our SKA-3yr parameter set. We have taken 15.87
points per decade in the model power spectrum and have used no
tomographic information. We assume $f_{\rm sky}=0.25$.}
\label{fig:pofk_SKA} 
\end{figure}

Two-dimensional $\hat{\kappa}(\ell)$ measurements can also be used to
estimate the power spectrum of matter fluctuations directly, if the
background cosmology (and thus the evolution with redshift of the
matter fluctuations) is assumed to be known.  To show this, we model
the matter power spectrum by linearly interpolating between points
that are evenly spaced in $\ln k$.  The $\ln P(k)$ values are treated
as the model parameters in this case.  Equation (\ref{eq:fisher}) can
then be used to find the expected uncertainties in the power spectrum
reconstruction.  For this latter calculation we assume purely linear
evolution with redshift and fix the shape of the
power spectrum to be that given by the CAMB
computer code \citep{2000ApJ...538..473L}.\footnote{http:/camb.info}

The uncertainties in these $P(k)$ measurements at different $k$ values
will be correlated, and their values will depend on the number of
points used to represent the power spectrum -- the more points the
better the $k$-space resolution, but the larger the noise per point.
Figures~\ref{fig:pofk_LOFAR} and \ref{fig:pofk_SKA} show forecast
uncertainties for two of our observational parameter sets.  The
matter power spectrum should be extremely well determined with 3 years
of SKA data, and using tomographic information would improve the
estimate still further.

The comoving scales which contribute most to 21~cm lensing can be seen
clearly in figures~\ref{fig:pofk_LOFAR} and \ref{fig:pofk_SKA}.  This
technique is complementary to CMB fluctuation measurements which probe
structure on scales with $10^{-4}< k < 0.1$
\citep{2007ApJS..170..377S}.  The scales probed by 21~cm lensing and
by galaxy clustering surveys overlap partly, but in the former case
the mass is probed directly with no bias uncertainties, and the
effective redshift of the measurement is much higher (see the next
section) so that uncertainties due to nonlinearities are also reduced.
Baryon acoustic oscillations can be seen in the power spectrum in
figures~\ref{fig:pofk_LOFAR} and \ref{fig:pofk_SKA} in the range
$k=0.04$ -- $0.4~\Mpc^{-1}$.  They would be well measured in the SKA-3yr case.  These
features would provide additional leverage when estimating
cosmological parameters, but we do not take this into account in
section~\ref{sec:cosm-param-1} below.

\subsection{Tomographic information}
\label{sec:tomogr-inform}

\begin{figure}
\rotatebox{90}{
\includegraphics[width=6.0cm]{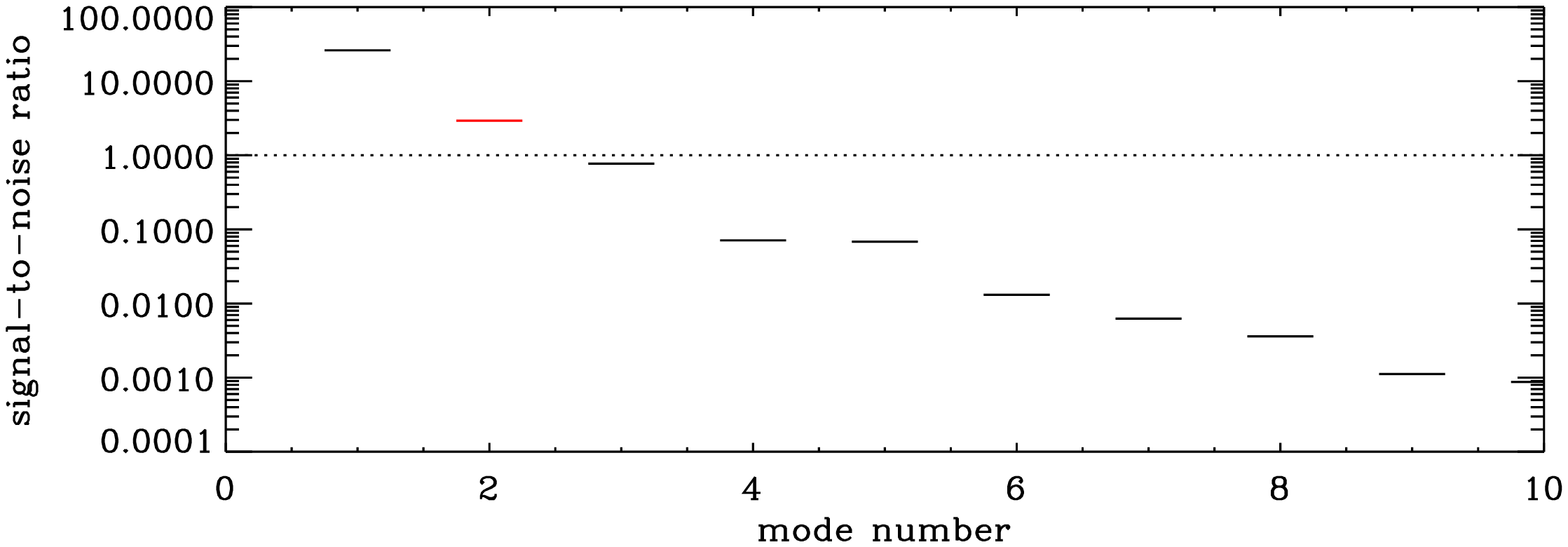}
\includegraphics[width=6.0cm]{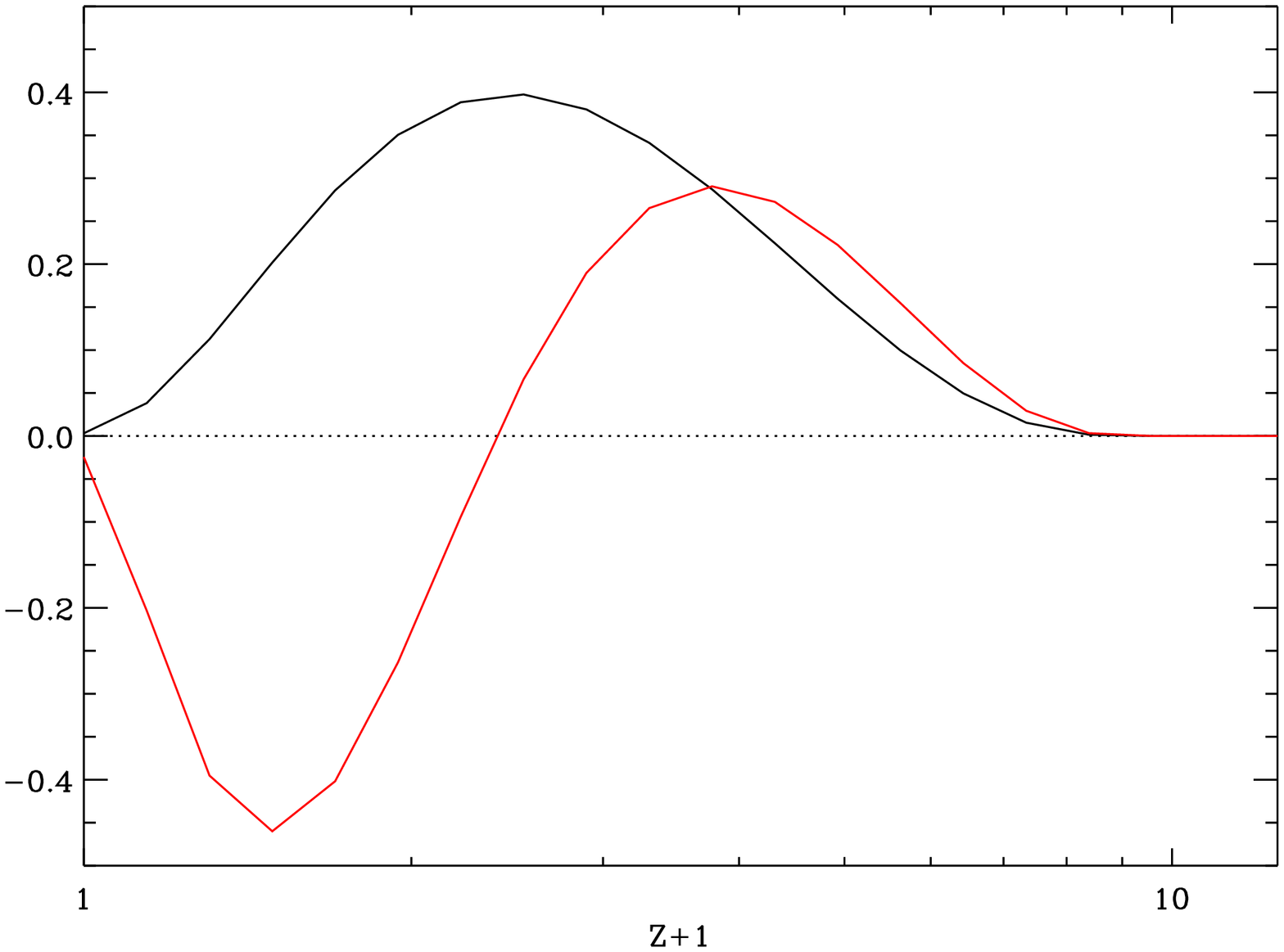}}  
\vspace{-2.5cm}
\caption{The uncorrelated modes in a decomposition of an estimate of
the structure growth function for the LOFARII-3yr case. The upper panel
shows the redshift dependence of the two modes which are expected to
yield estimates with signal-to-noise above 1, while the lower panel gives
the signal-to-noise values forecast for the 10 best constrained modes.
The $x$-axis is simply the mode number. Corresponding modes are indicated
using the same colour in each panel.  The growth function is
discretized into 20 points spaced evenly in $\log(z)$ and
$\kappa(\ell,z_s)$ are binned into 20 evenly spaced $z_s$ bins. We assume $f_{\rm sky}=0.25$.}
\label{fig:growth_stnLOFAR}
\end{figure}

\begin{figure}
\rotatebox{90}{
\includegraphics[width=6.0cm]{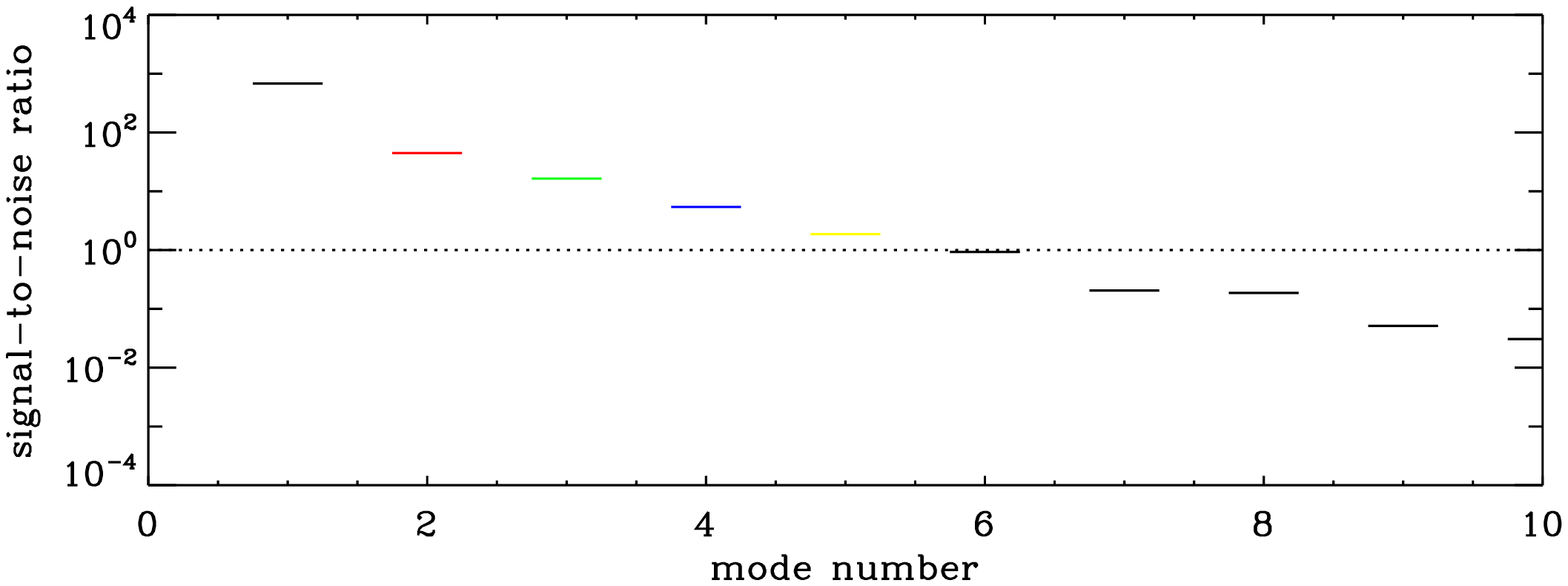}
\includegraphics[width=6.0cm]{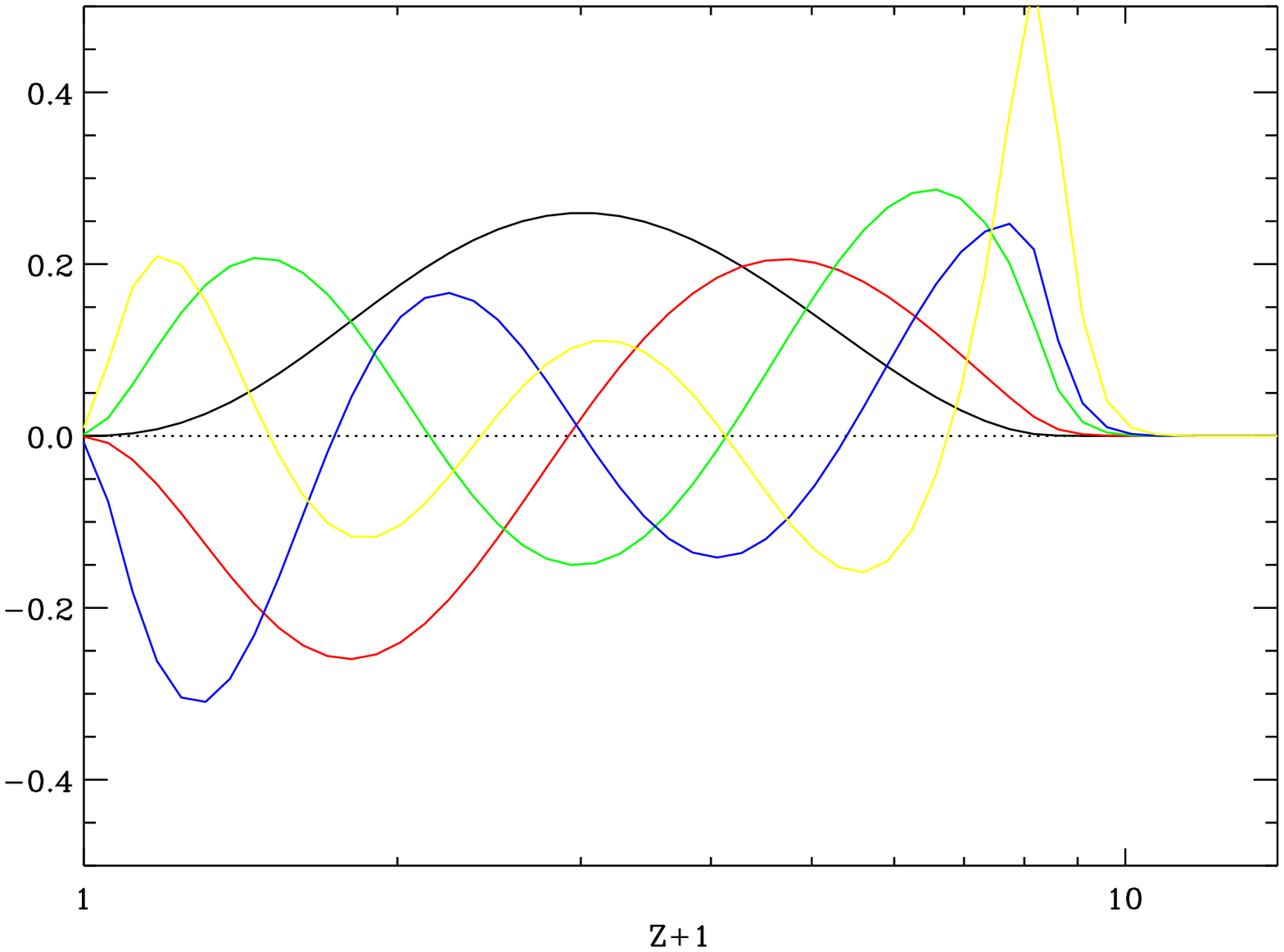}}  
\vspace{-2.5cm}
\caption{The uncorrelated modes in a decomposition of an estimate of
the structure growth function for the SKA-3yr case. The upper panel
shows the redshift dependence of the five modes which are expected to
yield estimates with signal-to-noise above 1, while the lower panel gives
the signal-to-noise values forecast for the 10 best constrained modes.  Corresponding modes are indicated using the same colour in each panel.  The growth function is
discretized into 50 points spaced evenly in $\log(z)$ and the
$\kappa(\ell,z_s)$ are binned into 20 evenly spaced $z_s$ bins.  We assume $f_{\rm sky}=0.25$.
}
\label{fig:growth_stnSKA} 
\end{figure}

As discussed in section~\ref{sec:param-estim-dens}, there is
tomographic information in the measured $\kappa(\ell,z_s)$.  This
makes it possible, in principle, to measure how $P(k)$ evolves with
redshift, and so to probe dark energy through its indirect effects on
the growth rate of linear fluctuations.

\begin{table*}
\begin{minipage}{17.5cm} 
\caption{Forecast constraints on cosmological parameters for full sky
surveys.  All constraints scale with sky coverage as $f_{\rm
sky}^{-1/2}$ except that the dark energy figure of merit, FOM, scales
as $f_{\rm sky}$.  The quantities in parentheses are the marginalized
uncertainties for constant $w$ ($w_a=0$). Both the high redshift 21~cm
emission and the lensed galaxy population are assumed to have been binned
into ten disjoint redshift intervals when making the corresponding
parameter estimates.}
\begin{tabular}{ll|ccccccc}
                           &                & $\ln
                           A$ & $n_s$ & $\Gamma$ & $\Omega_\Lambda$ & $w_o$& $w_a$ & FOM \\
                           & assumed values  & -9.6 ($\sigma_8=0.75$)
                           & 1  & 0.19 & 0.3 & -1.0 & 0 \\
\hline
\multirow{2}{*}{LOFARII-3yr} & marginalized  &  2.2 & 0.22 & 0.11 & 0.64 &
3.0 (1.3) & 11 & 0.075 \\
                           & unmarginalized   & 0.0047 & 0.0021 & 0.00038 & 0.0025 & 0.027 & 0.17  \\
\hline
LOFARII-3yr & marginalized & 0.036 & 0.0036 & 0.0012 & 0.0060 &  0.43 (0.053)
& 1.6 &  12 \\
+ Planck & unmarginalized & 0.0049 & 0.0016 & 0.00011 & 0.00047 &
0.0066 &  0.023 \\
\hline
LOFARII-3yr & marginalized & 0.032 & 0.0045 & 0.0017 & 0.0033 & 0.028
(0.0087) & 0.087 & 1300\\
+ galaxies & unmarginalized & 0.00032 & 0.00030 & $2.2\times 10^{-5}$
& 0.00015 & 0.0011 & 0.0053 \\
\hline
LOFARII-3yr & marginalized & 0.0086 & 0.0033 & 0.00053 & 0.0021 &
0.018 (0.0050) & 0.056 & 3600 \\
+ galaxies + Planck & unmarginalized & 0.00034 & 0.00030 & $2.1\times
10^{-5}$ & 0.00014 & 0.0011 &  0.0052 \\
\hline
\multirow{2}{*}{SKA-3yr} & marginalized & 0.029 & 0.011 & 0.0016 &
0.013 & 0.044 (0.026) & 0.12 &  310 \\
 & unmarginalized  & 0.00020 & 0.00023 & $1.3\times 10^{-5}$ & 0.00024
& 0.0018 & 0.011 \\
\hline
SKA-3yr & marginalized & 0.012 &   0.0036 &  0.00068 & 0.0027 & 0.036
(0.011) & 0.12 &  760 \\
+ Planck & unmarginalized & 0.00020 & 0.00023 & $1.3\times 10^{-5}$ &
0.00022 & 0.0017 & 0.0097 \\
\hline
SKA-3yr & marginalized & 0.017 & 0.0029 &  0.00096 & 0.0022 & 0.018 (0.0031) &
0.054 & 6000 \\
+ galaxies & unmarginalized & 0.00018 & 0.00019 & $1.1\times 10^{-5}$
& 0.00012 & 0.00083 & 0.0039 \\
\hline
SKA-3yr & marginalized & 0.0074 & 0.0027 &  0.00046 & 0.0017 & 0.014 (0.0020)  &  0.046 & 11000\\
+ galaxies + Planck & unmarginalized & 0.00018 &  0.00019 & $1.1\times
10^{-5}$ &  0.00012 &  0.00082 & 0.0039 \\
\hline
\multirow{2}{*}{galaxy lensing} & marginalized & 0.045 & 0.0052 &
0.0024 & 0.0036 &  0.030 (0.010) & 0.10 & 970 \\
& unmarginalized  & 0.00034 & 0.00030 & $2.2\times 10^{-5}$ & 0.00015
& 0.0011 &  0.0053 \\ 
\end{tabular}
\label{table:cosmo_params}
\end{minipage}
\end{table*}

While density perturbations are small they grow linearly,
$\delta(\boldsymbol{\ell},z) = D_g(z)
\delta(\boldsymbol{\ell},z=z_o)/D_g(z_o)$ where $z_o$ is some initial
redshift. Within General Relativity the linear growth function $D_g(z)$
is directly related to the cosmic expansion history $a(z)$ and so
through the Friedmann equation to the history of the various
contributions to the cosmic energy density. In other theories of
gravity $D_g(z)$ and $a(z)$ are independent
\citep{Peebles:80,2006ApJ...648..797B}.  A comparison of precise
measurements of the two can thus be used to test Einstein's theory.

Although structure growth is not precisely linear for all the $\ell$
values and for the full range of redshifts probed by 21~cm lensing, it
is probably a good approximation for the modes that are probed with high 
signal to noise.  
Without a specific theory for the nature and amount of dark energy the
exact form of $D_g(z)$ is unknown. In this section we model $D_g(z)$ as
piecewise linear between a set of interpolation redshifts at which we
treat its values as unknown parameters to be estimated. Note that
this implies more freedom than is physically allowed, since $D_g(z)$
should, for example, decrease smoothly and monotonically with
redshift. Technically our approach is not fully self-consistent since
the angular size distances in $\kappa(\vec{\ell},z_s)$ 
[equation~(\ref{eq:kappa_cont})] depend on the cosmic expansion history
and so should change when $D_g(z)$ changes.  Such effects are weaker
than those coming from the growth function itself, so we neglect them
here.  In section~\ref{sec:cosm-param-1} below, we investigate an
alternative approach in which we parameterize the equation of state of
dark energy and then use the Friedmann and linear growth equations to
calculate its consequences for $a(z)$ and $D_g(z)$. 

The Fisher matrix for the uncertainties in the estimated values of
$D_i=D_g(z_i)$ is highly non-diagonal, showing that these parameters
cannot all be estimated independently.  To understand what properties
of the growth function can actually be constrained, it is useful to
decompose $D_g(z)$ into modes that diagonalize the Fisher matrix, as in
equation~(\ref{eq:fisher_diag}).  Figures~\ref{fig:growth_stnLOFAR}
and \ref{fig:growth_stnSKA} show the well constrained modes in such
decompositions, together with forecasts of their signal-to-noise
ratios for surveys corresponding to our LOFARII-3yr and SKA-3yr cases.
We choose the $z_i$ to be evenly spaced in $\ln(1+z)$ because we find
this to give the best results with the fewest points, but we have
tried several other spacings and the results are, as expected, nearly
independent of this choice.

In the LOFARII-3yr case (figure~\ref{fig:growth_stnLOFAR}), estimates for
two modes are forecasted to have signal-to-noise above 2. The best
measured mode can be interpreted as the normalization of the power
spectrum at $z\sim 2$, while the higher order modes measure the rate
, acceleration and higher derivatives of structure growth over the redshift range, $z
\sim 0.5$ to 4.

In the SKA-3yr case (figure~\ref{fig:growth_stnSKA}), five modes should
have amplitude estimates with signal-to-noise greater than 1.  It should
thus be possible to map the growth function for $z\simgt 0.5$.  In
standard $\Lambda$CDM $D_g(z)$ is characterized by only two
parameters, $\Omega_{\rm matter}$ and $\Omega_\Lambda$, while four or five
could be measured in this way, providing a test of the consistency of
this simple model and an opportunity to discover something new.


The complete 3-dimensional density field could also, in principle, be
reconstructed.  Unlike the case of growth function reconstruction,
there is no longer any sample variance, since this part of the
``noise'' is now the signal.  The procedure for doing such
reconstructions is described in section~\ref{sec:tomography}, but we
will not investigate it further in this paper since it does not probe
the background cosmology directly.

\subsection{Cosmological parameters}
\label{sec:cosm-param-1}

Within any particular model for dark energy, the standard cosmological
parameters can be constrained directly, as described in
section~\ref{sec:cosm-param}.  We here investigate constraints on a
set of six parameters $\{\ln A,n_s,\Gamma,\Omega_\Lambda,w_o,w_a\}$.
$A$ is the dimensionless normalization of the primordial power spectrum and $n_s$ is its spectral index.  $\Gamma$ is the power
spectrum shape parameter \citep{1992MNRAS.258P...1E} given by $\Gamma
\simeq \Omega_mh \exp(-2\Omega_b)$ \citep{1994MNRAS.267.1020P}.  The first
three parameters only affect the shape and amplitude of the matter
power spectrum. The other three determine the expansion history $a(z)$
and the linear growth function $D_g(z)$.  In this parameterization the
lensing results are independent of $H_o$ (although the CMB constraints are
not) and, as a result, many of the parameter
degeneracies which can cause numerical instabilities when inverting
the Fisher matrix are avoided. We assume the Universe to be flat,
$\Omega_m+\Omega_\Lambda=1$.  This parameter set conforms to that used
by the Dark Energy Task Force
(DETF)\footnote{http://www.nsf.gov/mps/ast/detf.jsp} in their
investigation of the relative merits of different observational probes
of dark energy.  The dark energy equation of state parameter is here
assumed to evolve as $w(a) \equiv p/\rho = w_o + (1-a) w_a$, where
$a=(1+z)^{-1}$.  This is an arbitrary, but useful model that has
become popular in the literature.

\begin{table*}
\begin{minipage}{17.5cm}
\caption{As Table~\ref{table:cosmo_params} except that the sky coverage
  of the 21~cm lensing surveys is taken to be 25\% and that of the galaxy
  lensing survey to be 48\%.}
\begin{tabular}{ll|ccccccc}
     &    & $\ln A$ & $n_s$ & $\Gamma$ & $\Omega_\Lambda$ & $w_o$ & $w_a$ & FOM \\
& assumed values & -9.6 ($\sigma_8=0.75$) & 1 & 0.19 & 0.3 & -1.0 & 0 \\
\hline
LOFARII-3yr & marginalized & 0.052 & 0.0068 & 0.0028 & 0.0049 & 0.041 (0.013)
& 0.13 & 570\\
+ galaxies & unmarginalized & 0.00049 & 0.00043 & $3.2\times 10^{-5}$
& 0.00021 & 0.0016 & 0.0076 \\
\hline
SKA-3yr & marginalized & 0.029 & 0.0046 & 0.0016 & 0.0036 & 0.029 (0.0048) &
0.091 & 2300 \\
+ galaxies & unmarginalized & 0.00032 & 0.00032 & $2.0\times 10^{-5}$
& 0.00019 & 0.0013 & 0.0064 \\
\hline
SKA-3yr & marginalized & 0.0098 & 0.0035 &  0.00061 & 0.0022 & 0.020 (0.0028) &
0.064 & 5600 \\
+ galaxis + Planck & unmarginalized & 0.00032 & 0.00032 &
$2.0\times 10^{-5}$ & 0.00018 & 0.0013 &   0.0062 \\
\hline
galaxies & marginalized & 0.065 & 0.0075 & 0.0034 & 0.0053 & 0.043
(0.014) & 0.15 &  470 \\
 & unmarginalized & 0.00049 & 0.00044 & $3.2\times 10^{-5}$ & 0.00021
& 0.0016 & 0.0077 \\
\end{tabular}
\label{table:cosmo_params_cutsky}
\end{minipage}
\end{table*}

Table~\ref{table:cosmo_params} shows forecasts of the uncertainties in
estimates of these parameters both unmarginalized and marginalized
over all other parameters in the set.  The difference between the
marginalized and unmarginalized values indicates the importance of
parameter degeneracies. For ease of comparison, these forecasts
(unrealistically) assume full sky coverage for the 21~cm and galaxy
lensing surveys; the quoted
uncertainties scale as $f_{sky}^{-1/2}$ for lower sky coverage accept
when CMB constraints are included. The
first four parameters are constrained even in the LOFARII-3yr case,
but the same is not true for the parameters of the dark energy
equation of state. This reflects the fact that, in the model we are
assuming, dark energy is insignificant for the growth of structure at
$z\simgt 1$, while, as demonstrated in
section~\ref{sec:tomogr-inform}, 21~cm lensing is most sensitive to
structure growth above $z\sim 1$.

The DETF introduced a figure of merit (FOM) in order to compare the
power of different observational techniques to constrain dark energy.
The figure of merit is inversely proportional to the area of the error
ellipse in the $w_o$--$w_a$ plane after marginalizing over the other
parameters.\footnote{The figure of merit is normalized in different
ways by different authors, and even within the DETF report itself,
which defines it as the inverse of the area within the two sigma error
ellipse but usually quotes it as $4\pi$ times this, as we do here.
The absolute value has no particular significance so we will stick
with what seems to be the {\it de facto} DETF convention.}  These
values are given in the last column of Table~\ref{table:cosmo_params}.

The parameter degeneracies in 21~cm lensing can be significantly
reduced by incorporating information at low redshift from galaxy
lensing surveys.  The noise in power spectrum estimates from such
surveys can be written as $N_\kappa(\ell) = \sigma^2_\epsilon / n_{\rm
g}$ where $n_{\rm g}$ is the angular number density of background
galaxies and $\sigma_\epsilon$ is the root-mean-square intrinsic
ellipticity of those galaxies.  This neglects all systematic errors as
well as photometric redshift uncertainties. The latter can be
important for tomographic measurements.  Following standard
assumptions, we model the redshift distribution of usable galaxies as
$\eta(z) \propto z^2 e^{-(z/z_o)^{1.5}}$, where $z_o$ is set by the
desired median redshift, and we adopt $\sigma_\epsilon = 0.25$. The
proposed satellite SNAP\footnote{snap.lbl.gov} is expected to achieve
a usable galaxy density of $n_{\rm g}\simeq100\mbox{ arcmin}^{-2}$
with a median redshift $z\sim 1.23$, but it would only survey $\sim
2\%$ of the sky at this depth.  The
DUNE\footnote{www.dune-mission.net} mission proposes to survey almost
the whole extragalactic sky to a usable density of $n_{\rm g}\simeq
35\mbox{ arcmin}^{-2}$ with a median redshift of $z\sim 0.9$. Several
planned ground-based surveys -- LSST\footnote{www.lsst.org},
PanSTARRS\footnote{pan-stars.ifa.hawaii.edu},
VISTA\footnote{www.vista.ac.uk} will cover comparable areas to DUNE at
a similar depth.  Since the larger sky coverage more than outweighs
the reduced depth when constraining cosmological parameters, we adopt
the "DUNE" parameter set when comparing the power of 21~cm, galaxy, and
combined lensing surveys.  In order to use tomographic information, we
divide the galaxies into 10 redshift bins each containing the same
number of galaxies.  

We also include the predicted constraints from CMB observations with
the Planck Surveyor Satellite.  This is done by adding the Fisher
matrix for the lensing surveys to the Planck Fisher matrix as
calculated in \cite{2008arXiv0810.0003R}.  The CMB alone puts almost no constraint on $w_o$ and $w_a$, but the parameter degeneracies in the lensing surveys are greatly reduced by including CMB information.
Our lensing Fisher matrix computer code was tested against that of
\cite{2007arXiv0710.5171A} in order to
ensure the accuracy of the uncertainties we forecast for parameter
estimates.

Table~\ref{table:cosmo_params} compares the uncertainties forecast for a 21~cm
lensing survey alone with those forecast for the galaxy lensing survey alone, and for combinations of surveys.  A full-sky 21~cm survey
with LOFARII-3yr parameters would not constrain these parameters as
well as a full-sky "DUNE" survey.  The full-sky SKA-3yr survey would
produce constraints that are similar to the galaxy survey on this
parameter set.  As noted above, all the action occurs at
relatively low redshift for this particular dark energy model.
When the two types of surveys are combined they give much tighter 
constraints than either survey alone.  Even the seemingly insensitive
LOFARII-like observations can improve the dark energy constraints (a
factor of 1.3 in the FOM) when combined with the galaxy survey. Combining a survey with SKA-like sensitivity with a galaxy lensing survey improves the FOM by a factor of $\sim 6$ and when CMB data is included the improvement is more than on order of magnitude.

In practice, 21~cm lensing surveys are not likely to be full-sky,
whereas at least half the sky could be covered by a satellite-based
galaxy lensing survey (regions at low Galactic latitude are unusable).
In Table~\ref{table:cosmo_params_cutsky} we combine 21~cm lensing
surveys covering 25\% of the sky with a galaxy lensing survey covering
half the sky and the Planck CMB constraints.  Even with this reduced size the combination of the two
surveys results in a drastic improvement in the dark energy FOM and in
other parameter constraints when compared with any of the surveys by itself.

It is important to recognize that the constraints on dark energy found
here for 21~cm surveys alone are in large part a consequence of the
particular parameterization adopted for the dark energy equation of
state.  It could be that the energy density of dark energy is still
significant above $z\sim 1$ or that the rate of structure formation is
otherwise significantly modified at such redshifts.  This is the case in
many more physically motivated models, for example, early dark energy
models \citep{1988PhRvD..37.3406R,1998PhRvL..80.1582C}, 4-D gravity
\citep{2000PhLB..485..208D}, coupled scalar field theories
\citep{2000MNRAS.312..521A} and some possible modifications of the
underlying theory of gravity \citep[for example]{2004PhRvD..70d3528C}.
In some of these models the growth factor could differ from the
standard one by a factor of 2 at $z=2$.  Galaxy lensing alone would
not able to distinguish between many of these theories, because of the
poor leverage it offers on the high redshift Universe.  As shown in
section~\ref{sec:tomogr-inform}, 21~cm lensing could probe the energy
density evolution back to $z\simeq 7$.

\section{conclusions}
\label{sec:discussion}

We have shown that a very large amount of cosmological information
could be extracted from the gravitational lensing of pregalactic 21~cm
radiation.  The planned low-frequency radio telescopes, an upgraded LOFAR or SKA,
will already have sufficient sensitivity to do this successfully as
long as reionization occurs sufficiently late, foreground
contamination can be accurately removed, and enough sky can be
observed. Under these conditions, the matter power spectrum should be
well measured in two and three dimensions, and the growth of structure
will be probed from $z=0.5$ to $z=7$.  Such observations would
constrain many cosmological quantities, in particular the form of the
primordial power spectrum and presence of early dark energy, to
unprecedented accuracy.

Combining 21~cm lensing surveys with surveys of lensing of foreground
galaxies would dramatically improve constraints on structure growth at
$z\simlt 1$ where the strongest effects occur in the most popular
parameterizations of dark energy.  Increasing the collecting area
of the telescopes by a factor of $\sim 2$ would also dramatically increase
their sensitivity to lensing, as would extending their frequency range to
lower values, if foregrounds permit.

\vspace{0.3cm} 
\leftline{\bf Acknowledgments} 
We would like to thank Adam Amara for allowing use of his own software
to compare with results from our own parameter constraint codes.  We would like than J. Weller for help in calculating the expected fisher matrix for the Planck satellite mission.


 \bibliographystyle{/Users/bmetcalf/Work/TeX/apj/apj}
 \bibliography{/Users/bmetcalf/Work/mybib}

\begin{thebibliography}{28}
\expandafter\ifx\csname natexlab\endcsname\relax\def\natexlab#1{#1}\fi

\bibitem[{{Amara} \& {Refregier}(2007)}]{2007arXiv0710.5171A}
{Amara}, A. \& {Refregier}, A. 2007, ArXiv e-prints, 710

\bibitem[{{Amendola}(2000)}]{2000MNRAS.312..521A}
{Amendola}, L. 2000, \mnras, 312, 521

\bibitem[{{Bertschinger}(2006)}]{2006ApJ...648..797B}
{Bertschinger}, E. 2006, \apj, 648, 797

\bibitem[{{Caldwell} {et~al.}(1998){Caldwell}, {Dave}, \&
  {Steinhardt}}]{1998PhRvL..80.1582C}
{Caldwell}, R.~R., {Dave}, R., \& {Steinhardt}, P.~J. 1998, Physical Review
  Letters, 80, 1582

\bibitem[{{Carroll} {et~al.}(2004){Carroll}, {Duvvuri}, {Trodden}, \&
  {Turner}}]{2004PhRvD..70d3528C}
{Carroll}, S.~M., {Duvvuri}, V., {Trodden}, M., \& {Turner}, M.~S. 2004, \prd,
  70, 043528

\bibitem[{{Dvali} {et~al.}(2000){Dvali}, {Gabadadze}, \&
  {Porrati}}]{2000PhLB..485..208D}
{Dvali}, G., {Gabadadze}, G., \& {Porrati}, M. 2000, Physics Letters B, 485,
  208

\bibitem[{{Efstathiou} {et~al.}(1992){Efstathiou}, {Bond}, \&
  {White}}]{1992MNRAS.258P...1E}
{Efstathiou}, G., {Bond}, J.~R., \& {White}, S.~D.~M. 1992, \mnras, 258, 1P

\bibitem[{{Field}(1959)}]{1959ApJ...129..536F}
{Field}, G.~B. 1959, \apj, 129, 536

\bibitem[{{Furlanetto} {et~al.}(2006){Furlanetto}, {Oh}, \&
  {Briggs}}]{astro-ph/0608032}
{Furlanetto}, S.~R., {Oh}, S.~P., \& {Briggs}, F.~H. 2006, \physrep, 433, 181

\bibitem[{{Hilbert} {et~al.}(2007){Hilbert}, {Metcalf}, \& {White}}]{HMW07}
{Hilbert}, S., {Metcalf}, R.~B., \& {White}, S.~D.~M. 2007, \mnras, 382, 1494

\bibitem[{{Hu} \& {Keeton}(2002)}]{2002PhRvD..66f3506H}
{Hu}, W. \& {Keeton}, C.~R. 2002, \prd, 66, 063506

\bibitem[{{Hu} \& {Okamoto}(2002)}]{2002ApJ...574..566H}
{Hu}, W. \& {Okamoto}, T. 2002, \apj, 574, 566

\bibitem[{{Lewis} {et~al.}(2000){Lewis}, {Challinor}, \&
  {Lasenby}}]{2000ApJ...538..473L}
{Lewis}, A., {Challinor}, A., \& {Lasenby}, A. 2000, \apj, 538, 473

\bibitem[{{Lu} \& {Pen}(2007)}]{2007arXiv0710.1108L}
{Lu}, T. \& {Pen}, U.-L. 2007, ArXiv e-prints, 710

\bibitem[{{Madau} {et~al.}(1997){Madau}, {Meiksin}, \&
  {Rees}}]{1997ApJ...475..429M}
{Madau}, P., {Meiksin}, A., \& {Rees}, M.~J. 1997, \apj, 475, 429

\bibitem[{{McQuinn} {et~al.}(2006){McQuinn}, {Zahn}, {Zaldarriaga},
  {Hernquist}, \& {Furlanetto}}]{2006ApJ...653..815M}
{McQuinn}, M., {Zahn}, O., {Zaldarriaga}, M., {Hernquist}, L., \& {Furlanetto},
  S.~R. 2006, \apj, 653, 815

\bibitem[{{Metcalf} \& {White}(2007)}]{metcalf&white2006}
{Metcalf}, R.~B. \& {White}, S.~D.~M. 2007, \mnras, 381, 447

\bibitem[{{Morales}(2005)}]{2005ApJ...619..678M}
{Morales}, M.~F. 2005, \apj, 619, 678

\bibitem[{{Peacock} \& {Dodds}(1994)}]{1994MNRAS.267.1020P}
{Peacock}, J.~A. \& {Dodds}, S.~J. 1994, \mnras, 267, 1020

\bibitem[{{Peacock} \& {Dodds}(1996)}]{peac96}
---. 1996, \mnras, 280, L19

\bibitem[{Peebles(1980)}]{Peebles:80}
Peebles, P.~J.~E. 1980, The Large-Scale Structure of the Universe (Princeton,
  NJ: Princeton University Press)

\bibitem[{{Rassat} {et~al.}(2008){Rassat}, {Amara}, {Amendola}, {Castander},
  {Kitching}, {Kunz}, {Refregier}, {Wang}, \& {Weller}}]{2008arXiv0810.0003R}
{Rassat}, A., {Amara}, A., {Amendola}, L., {Castander}, F.~J., {Kitching}, T.,
  {Kunz}, M., {Refregier}, A., {Wang}, Y., \& {Weller}, J. 2008, ArXiv e-prints

\bibitem[{{Ratra} \& {Peebles}(1988)}]{1988PhRvD..37.3406R}
{Ratra}, B. \& {Peebles}, P.~J.~E. 1988, \prd, 37, 3406

\bibitem[{{Spergel} {et~al.}(2007){Spergel}, {Bean}, {Dor{\'e}}, {Nolta},
  {Bennett}, {Dunkley}, {Hinshaw}, {Jarosik}, {Komatsu}, {Page}, {Peiris},
  {Verde}, {Halpern}, {Hill}, {Kogut}, {Limon}, {Meyer}, {Odegard}, {Tucker},
  {Weiland}, {Wollack}, \& {Wright}}]{2007ApJS..170..377S}
{Spergel}, D.~N., {Bean}, R., {Dor{\'e}}, O., {Nolta}, M.~R., {Bennett}, C.~L.,
  {Dunkley}, J., {Hinshaw}, G., {Jarosik}, N., {Komatsu}, E., {Page}, L.,
  {Peiris}, H.~V., {Verde}, L., {Halpern}, M., {Hill}, R.~S., {Kogut}, A.,
  {Limon}, M., {Meyer}, S.~S., {Odegard}, N., {Tucker}, G.~S., {Weiland},
  J.~L., {Wollack}, E., \& {Wright}, E.~L. 2007, \apjs, 170, 377

\bibitem[{{Takada} \& {Jain}(2004)}]{2004MNRAS.348..897T}
{Takada}, M. \& {Jain}, B. 2004, \mnras, 348, 897

\bibitem[{{Vale} \& {White}(2003)}]{2003ApJ...592..699V}
{Vale}, C. \& {White}, M. 2003, \apj, 592, 699

\bibitem[{{Zahn} \& {Zaldarriaga}(2006)}]{ZandZ2006}
{Zahn}, O. \& {Zaldarriaga}, M. 2006, \apj, 653, 922

\bibitem[{{Zaldarriaga} {et~al.}(2004){Zaldarriaga}, {Furlanetto}, \&
  {Hernquist}}]{2004ApJ...608..622Z}
{Zaldarriaga}, M., {Furlanetto}, S.~R., \& {Hernquist}, L. 2004, \apj, 608, 622

\end{thebibliography}

\end{document}